\documentclass[preprint,showpacs,amsmath,amssymb,prl,superscriptaddress]{revtex4}

\usepackage{graphicx}
\usepackage{dcolumn}
\usepackage{bm}

\begin{document}


\title{Spontaneous emergence of angular momentum Josephson oscillations\\ in coupled annular Bose-Einstein condensates}
\date{\today}

\author{Igor Lesanovsky}
\email[]{igor@iesl.forth.gr}
\affiliation{%
Institute of Electronic Structure and Laser,
Foundation for Research and Technology - Hellas, P.O. Box 1527, GR-71110 Heraklion, Greece}%
\author{Wolf von Klitzing}
\email[]{wvk@iesl.forth.gr}
\affiliation{%
Institute of Electronic Structure and Laser, Foundation for Research
and Technology - Hellas, P.O. Box 1527, GR-71110 Heraklion, Greece}
\date{\today}

\begin{abstract}\label{txt:abstract}
We investigate the nonlinear dynamics of two coupled annular
Bose-Einstein condensates (BECs). For certain values of the coupling
strength the nonrotating ground state is unstable with respect to
fluctuations in the higher angular momentum modes. The two branched
Bogoliubov spectrum exhibits distinct regions of instability
enabling one to selectively occupy certain angular momentum modes.
For sufficiently long evolution times angular momentum Josephson
oscillations spontaneously appear, breaking the initial chiral
symmetry of the BECs.
\end{abstract}
\maketitle

One of the most famous paradigms of quantum physics is the existence
of Josephson oscillations. They were first predicted for
superconductors separated by an insulating layer
\cite{Josephson62PL}. Later they have been observed in superfluid
$^3\text{He}$ \cite{Davis02} and gaseous BECs
\cite{Smerzi97,Albietz05}. Next to the oscillations of charge and/or
particles between two modes these systems can exhibit highly
nonlinear dynamics with sometimes surprising behavior. In this
letter we study one-dimensional (1D) BECs confined in two
ring-shaped traps which are sufficiently close to each other to
allow tunneling through the barrier between them. We demonstrate
that the stationary state, in which only the zero angular mode is
occupied by the BECs, becomes unstable for certain values of the
coupling strength. This allows for a selective occupation of
specific angular momentum modes. For short propagation times the
angular momentum in each ring is conserved. For longer interaction
times, however, angular momentum Josephson oscillations appear. This
novel type of Josephson oscillations spontaneously breaks the
initial chiral symmetry of the individual BECs.

In the mean-field description the evolution of a dilute gas of
identical interacting Bosons under the influence of the trapping
potential $V(\mathbf{r})$ is governed by the Gross-Pitaevskii
equation (GPE), which in cylindrical coordinates reads
\begin{eqnarray}
  i\hbar\partial_t
  \Psi=[\frac{\hbar^2}{2M}(-\partial^2_\rho-\partial_z^2+\frac{L_z^2}{\hbar^2\rho^2})+V(\mathbf{r})]\Psi+g|\Psi|^2\Psi
  \label{eq:gpe}
\end{eqnarray}
where $M$ is the atomic mass, $g$ the nonlinear coupling constant,
$\Psi=\Psi(\mathbf{r})$ the bosonic mean field and
$L_z=-\hbar^2\partial_\phi^2$ the $z$-component of the angular
momentum operator. The system consist of two BECs in parallel
ring-shaped traps encircling the $z$-axis. The positions of the
upper and lower ring are $\pm z_0$, respectively. Correspondingly,
the trapping potential takes the form
$V(\mathbf{r})=V_\rho(\rho-\rho_0)+V_z(z,z_0)$. The first term
provides harmonic radial confinement centered at $\rho=\rho_0$, and
$V_z(z,z_0)$ creates a symmetric double well potential with its
minima at $z=\pm z_0$. Both BECs reside in the radial ground state
$\Psi_\rho(\rho)$ of $V_\rho(\rho-\rho_0)$. Vertically they occupy
the harmonic ground states $\Phi(z\pm z_0)$ which are localized in
the upper and the lower well of $V_z(z,z_0)$, respectively. The
total wave function of the system can then be written as
$\Psi(\mathbf{r})=\Psi_\rho(\rho)\left[\Phi(z-z_0)\chi_u(\phi)+\Phi(z+z_0)\chi_d(\phi)\right]$
where the indices $u$ and $d$ refer to the upper and lower ring.
After inserting $\Psi(\mathbf{r})$ into the GPE we obtain the two
coupled equations
\begin{eqnarray}
  i\partial_\tau\chi_{u/d}&=&-\partial_\phi^2\chi_{u/d}+\kappa \chi_{d/u}+\gamma|\chi_{u/d}|^2\chi_{u/d}\label{eq:left_ring}
\end{eqnarray}
Here we have introduced the scaled time $\tau=\frac{\hbar}{2M
R^2}t$, the coupling $\kappa=-R^2\int
dz\,\Phi(z+z_0)\left[\partial_z^2-\frac{2M}{\hbar^2}V(z)\right]\Phi(z-z_0)$
and the interatomic interaction parameter $\gamma=\frac{2M R^2\,
g_{1D}}{\hbar^2}\int dz\,\Phi^4(z)$ with $R^{-2}=\int d\rho\,
\rho^{-2} |\Psi_\rho(\rho)|^2$ \footnote{Unless otherwise stated all
quantities will be given in units of the scaled time
$\tau_0=\frac{2M R^2}{\hbar}$ and energy $E_0=\frac{\hbar^2}{2M
R^2}$}. We assume that the external confinement allows an effective
1D treatment of the BECs. The atom-atom interaction can then be
described by the 1D coupling constant
$g_{1D}=\frac{2\hbar^2}{M}\frac{a}{a_\rho^2}$ \cite{Das02}. Here $a$
is the three-dimensional s-wave scattering length and $a_\rho$ the
harmonic oscillator length of the radial ground state. Equations,
similar to eq.(\ref{eq:left_ring}) arise in the context of two
coupled elongated condensates \cite{Bouchoule05}. Ring traps
additionally allow the existence of stationary currents.

In the angular momentum mode representation the azimuthal wave
function of the individual BECs can be written according to
$\chi_{u/d}=(2\pi)^{-1/2}\exp(i\,\theta_{u/d})\sum_m \alpha^{u/d}_m
\exp(im \phi)$ with $\theta_{u/d}$ being the phase of the wave
function in the respective annulus. The coefficients
$\alpha^{u/d}_m$ are normalized such that
$|\alpha^{u/d}_m|^2=N^{u/d}_m$ corresponds to the number of
particles residing in the $m$-th angular momentum mode. Hence $\int
d\phi\, |\chi_{u/d}|^2=N^{u/d}$ corresponds to the total number of
particles in each of the two annuli. Inserting the above expression
for $\chi_{u/d}$ into eq.(\ref{eq:left_ring}) we find the system of
coupled equations
\begin{eqnarray}
  i\partial_\tau\alpha^{u/d}_m&=&m^2\alpha^{u/d}_m+\kappa_{u/d}
  \alpha^{d/u}_m+\frac{\gamma}{2\pi}\sum_{nn^\prime}\alpha^{u/d}_n {\alpha^*}^{u/d}_{n^\prime}
  \alpha^{u/d}_m \label{eq:coupled_m_equations}
\end{eqnarray}
with $\kappa_{u/d}=\kappa^*_{d/u}=\kappa\,
e^{i(\theta_d-\theta_u)}$. We now seek a stationary solution of this
system for which in both of the annuli solely the $m=0$ mode is
occupied. This exists only for equal number of particles, i.e.
$N^u_0=N^d_0=N_0$, and equal coupling $\kappa_u=\kappa_d=\kappa$. We
then find the two solutions
\begin{eqnarray}
\alpha_0^u=\pm\alpha_0^d=\sqrt{N_0}e^{i(\varepsilon \pm
\kappa)\tau+i\theta}&,&\alpha^{d/u}_{m\neq 0}=0
\label{eq:alpha_stationary}
\end{eqnarray}
with some arbitrary phase $\theta$ and $\varepsilon=\frac{\gamma
N_0}{2\pi}$ being the nonlinear energy due to the interatomic
interaction. Hence the total two-dimensional wave function becomes
either a symmetric or an anti-symmetric superposition of the axial
ground states of the two annuli:
$$\Psi_\pm(\mathbf{r})=\sqrt{N_0}e^{i\mu_\pm\tau+i\theta}\Psi_\rho(\rho)\left[\Phi(z+z_0)\pm\Phi(z-z_0)\right]$$
with $\mu_\pm=\varepsilon \pm \kappa$ being the chemical potential.

In order to investigate the stability of these states with respect
to fluctuations in modes with $m\neq 0$ we make the ansatz
$\alpha_{m\neq 0}^{u/d}=e^{i\mu_\pm \tau}\left[u_m^{u/d}e^{-i\omega
\tau}+{v^*}_m^{u/d}e^{i\omega \tau}\right]$. Inserting this together
with eq.(\ref{eq:alpha_stationary}) into the
eq.(\ref{eq:coupled_m_equations}) yields after linearization in the
$u_m$ and $v_m$ the eigenvalue equation
\begin{eqnarray}
  \omega u_m^{u/d}&=&(m^2+\mu_\pm) u_m^{u/d}+\varepsilon
  e^{i2\theta} v_{-m}^{u/d}+\kappa u_m^{d/u} \label{eq:bogoliubov_eq}\\
  -\omega v_{-m}^{u/d}&=&(m^2+\mu_\pm) v_{-m}^{u/d}+\varepsilon
  e^{-i2\theta} u_{m}^{u/d}+\kappa v_{-m}^{d/u}\nonumber.
\end{eqnarray}
We then find the excitation spectrum consisting of the two branches
\begin{eqnarray}
  \omega_\pm=\sqrt{(m^2+\varepsilon-\kappa
  \pm\kappa)^2-\varepsilon^2}.\label{eq:omega}
\end{eqnarray}
This result is obtained for either chemical potential $\mu_\pm$. The
branch $\omega_+$ corresponds to the well-known Bogoliubov spectrum
\cite{Pethick02} of a uniform BEC but with integer $m$. In this
letter we consider only repulsive interatomic interaction, i.e.
$\varepsilon\geq 0$. Thus $\omega_+$ is always a real number and the
coupled condensates are stable against fluctuations in this
excitation branch. Note that $\omega_+$ is independent of the
coupling strength $\kappa$. The $\omega_-$ branch on the other hand
depends on $\kappa$ and result from the interaction among the
coupled condensates. Its frequencies are either real or imaginary
but never complex. For zero coupling $\omega_-$ is identical with
$\omega_+$.
\begin{figure}[htb]
\center
\includegraphics[angle=0,width=8.5cm]{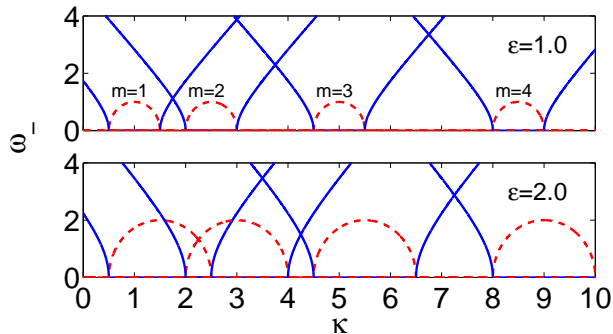}
\caption{(Color online) Branch $\omega_-$ of the Bogoliubov spectrum
of the coupled condensates plotted against the coupling strength
$\kappa$ for two values of the nonlinear energy $\varepsilon$
(blue/solid: real part, red/dashed: imaginary part). In panel a the
instable regions are labelled with the respective $m$ number. Due to
the symmetry of eq.(\ref{eq:bogoliubov_eq}) the spectrum is
symmetric with respect to a sign change of $m$.
}\label{fig:instability}
\end{figure}
Figure \ref{fig:instability} depicts for two values of the nonlinear
energy ($\varepsilon=1.0$ and $\varepsilon=2.0$) the spectrum
$\omega_-$ as a function of the coupling $\kappa$. The blue/solid
curves represent the real part and the red/dashed curves the
imaginary part of $\omega_-$. First, let us consider relatively low
energies $(\varepsilon < 2)$ (fig.$\,$\ref{fig:instability}a).
Starting from $\kappa=0$, we notice that $\omega_{-}$ is positive
and real, and that it decreases monotonously to $\omega_{-}=0$ at
$\kappa=1/2$. After this, one enters a region where $\omega_-$ is
imaginary with $\text{Im}(\omega_-)>0$. The system is then unstable
under fluctuations in the $m=\pm 1$ modes which grow at a rate of
$\Gamma=2\,\text{Im}(\omega_-)$ \cite{Bouchoule05,Saito06}. As
$\kappa$ increases regions of stability and instability follow one
another. The latter are defined by $m^2/2<\kappa<m^2/2+\varepsilon$.
For any given $\varepsilon$ the maximum growth rate
$\Gamma_\text{max}=2\varepsilon$ is a universal quantity for all
modes which is independent of $m$ and is established at the coupling
strengths $\kappa=\frac{1}{2}[m^2+\varepsilon]$.  Note that for
sufficiently large $\varepsilon$ the unstable regions of two
adjacent $m$-modes may even overlap thus eliminating the stable
region in between. This can be seen in
fig.$\,$\ref{fig:instability}b, where for a nonlinear energy
$\varepsilon=2.0$ part of the regions of instability for the $m=1$
and $m=2$ modes overlap. Since the coupling $\kappa$ is a function
of the trapping potential, and as such experimentally accessible,
the instability of the modes can be used to selectively affect one
or more $m$-modes of the rings. For example, at $\varepsilon=1.0$
each mode is "individually addressable" through an appropriate
choice of $\kappa$ (see fig.$\,$\ref{fig:instability}a). At
$\varepsilon=2.0$ and $\kappa=2.1$, on the other hand, both the
$m=1$ and $m=2$ are unstable with respect to fluctuations.
\begin{figure}[htb]
\center
\includegraphics[angle=0,width=9cm]{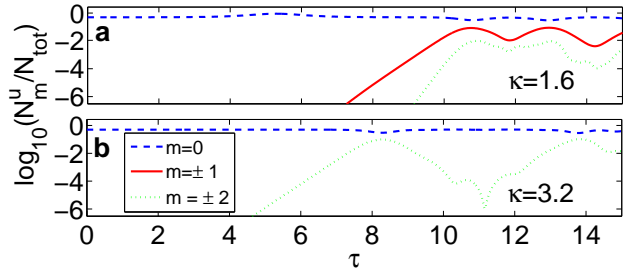}
\caption{(Color online) Evolution of the occupation number $N^u_m$
normalized to the total number of particles for the nonlinear energy
$\varepsilon=2.0$. For $\kappa=1.6$ (panel a) we observe an
exponential population increase in the $m=\pm 1$ modes. Population
of the $m=\pm 2$ mode is also visible at later times but is
suppressed by more than two orders of magnitude. For $\kappa=3.2$
(panel b), the accumulation in population of the $m=\pm 1$ modes is
suppressed and only the exponential growth of the $m=\pm 2$ modes is
visible.}\label{fig:selective_excitation}
\end{figure}

The growth of the instable modes eventually leads to a break down of
the linearized eqs.(\ref{eq:bogoliubov_eq}) due to the interaction
between higher lying $m$-modes. We therefore return to
eq.(\ref{eq:coupled_m_equations}) and integrate them numerically. We
choose $\alpha_0^{u/d}=\sqrt{N_0+\delta_{u/d}}$ as initial
condition, i.e. almost equal number of atoms in the $m=0$ mode of
both rings, whilst allowing for experimentally unavoidable particle
number fluctuations of the order of
$\delta_{u,d}=\mathcal{O}(\sqrt{N_0})$.  Adding these fluctuations
has only minor influence on the numerical results. Following Saito
\emph{et al.} \cite{Saito06}, we introduce a small seed in the
lowest few angular momentum modes (up to $m=\pm 5$) with a magnitude
of $10^{-4}\times \sqrt{N_0}$. Again, such a fluctuation is
experimentally inevitable. We truncate the set of coupled
eqs.(\ref{eq:coupled_m_equations}) at the angular momentum mode
$m=\pm 15$, well above the highest contributing mode. We verified
the quality of the propagation by monitoring energy, norm, and
angular momentum conservation. Since both annuli have a slight
population difference the nonlinear energy is calculated according
to $\varepsilon=\frac{\gamma N_\text{tot}}{4 \pi}$ with
$N_\text{tot}=N_u+N_d$. An example of the numerical propagation can
be seen in fig.$\,$\ref{fig:selective_excitation}. We show the
occupation $N^u_m$ for $m=0,\pm 1, \pm 2$ at the nonlinear energy
$\varepsilon=2.0$ (see also fig.$\,$\ref{fig:instability}b) for two
different coupling strengths $\kappa$. For $\kappa=1.6$ only the
$m=\pm 1$ modes are unstable. We observe for early times ($\tau<10$)
the predicted exponential increase in population with a rate of
$\Gamma\approx 4$. At later times $\tau>9$ the population also in
the $m=\pm 2$ modes increases slightly. This cannot be described by
the linearized eqs.(\ref{eq:bogoliubov_eq}). However, in the time
window considered here, the population of the $m=\pm 1$ modes is
more than two orders of magnitude larger than one of the $m=\pm 2$
modes. In the lower panel we present the same plot for $\kappa=3.2$.
Here we observe no population growth within the $m=\pm 1$ modes but
$N^u_{\pm 2}$ grows at a rate of $\Gamma=3.9$.  This clearly
demonstrates the possibility of a \emph{selective} angular momentum
mode excitation by tuning $\kappa$.

Let us now turn to the angular momentum of the two BECs. The $L_z$
expectation value of $\omega_-$ branch is
$\left<L_z^u\right>_-^{m}=\left<L_z^d\right>_-^{m}=0$ which implies
an equal population of states with opposite $m$. Conversely, for
modes of the branch $\omega_+$ we find
$\left<L_z^u\right>_+=\left<L_z^d\right>_+=\frac{m}{2}\frac{\sqrt{m^4+2\varepsilon
m^2}}{m^2+\varepsilon}$. However, according to eq.(\ref{eq:omega})
the value of $\omega_+$ is always real, which results in a growth
rate of $\Gamma=0$.
\begin{figure}[htb]
\center
\includegraphics[angle=0,width=8.8cm]{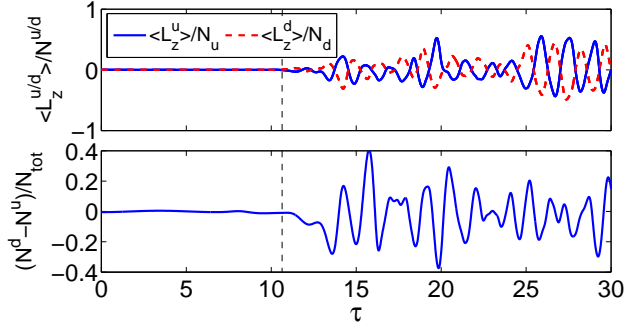}
\caption{(Color online) Angular momentum per particle in units of
$\hbar$ and relative particle difference between the two annuli
($\varepsilon=2.0$ and $\kappa=2.1$). For $\tau<11$ small
oscillations in the particle difference in the order of $10^{-2}$
take place. At $\tau=\tau_{osc}\approx 11$ (dashed vertical line) we
observe the onset of oscillations with bigger amplitude which are
accompanied by angular momentum Josephson oscillations.}
\label{fig:l_oscillations}
\end{figure}
In fig.$\,$\ref{fig:l_oscillations} we present the angular momentum
per particle and the relative particle number difference between the
coupled BECs. The nonlinear energy is again $\varepsilon=2.0$. The
coupling is $\kappa=2.1$, where both the $m=\pm 1$ and the $m= \pm
2$ are unstable. Until $\tau=\tau_\text{osc}\approx 11$ we find very
small oscillations of the relative particle difference which are of
the order of $10^{-2}$. The decompositions into $m$ modes shows that
this is due to small oscillations taking place between the $m=0$
modes of the two rings (see also
fig.$\,$\ref{fig:selective_excitation} for $\tau<6$). These are
``ordinary" Josephson oscillations. Here no formation of currents,
i.e $\left<L_z^{u/d}\right>\neq 0$, takes place in either of the
annuli. For $\tau>11$ the situation changes dramatically. We find
particle oscillations up to $|N^d-N^u|/N_\text{tot}=0.4$ and a
nonzero expectation values for $\left<L_z^{u/d}\right>$. The time
$\tau_\text{osc}$ of the onset of this regime depends on the
magnitude of the initial seed $\alpha_m$. A rough estimate for
$\tau_\text{osc}$ can be obtained from the equation $N_0 \approx
N_m=|\alpha_m(t=0)|^2\exp(\Gamma\,\tau_\text{osc})$. These
oscillations in $\left<L_z^{u/d}\right>$ are due to the population
of $\omega_+$ modes caused by the nonlinear evolution of the system.

The oscillations spontaneously emerging for $\tau>\tau_\text{osc}$
are Josephson oscillations of the angular momentum. They break the
chiral symmetry of the initial state's wave function where none of
the rings carried a net angular momentum.

Finally, we turn to the experimental realizability of the
``ordinary" and angular momentum Josephson oscillations. For
$^\text{87}\text{Rb}$ and a ring radius of $\rho_0=1.2\, \mu m$ the
energy scale given by the length of the ring evaluates to
$E_0\approx k_B \times 2\,nK$ and consequently we find a time scale
of $\tau_0\approx 4\, ms$.  For a radial oscillator length of
$a_\rho=0.3\,\mu m$ a nonlinear energy of $\varepsilon=2.0$ is
achieved for a particle number $N_0\approx 50$ \footnote{These
values have been calculated under the assumption that the states
$\Phi(z\pm z_0)$ are harmonic oscillator ground states with an
oscillator length $a_z=0.5\,\mu m$. The scattering length for
$^\text{87}\text{Rb}$ was taken to be $5.2\, nm$.}. The experimental
feasibility of building ring-shaped traps has been demonstrated
recently \cite{Gupta05,Hofferberth06}. We therefore hope the results
presented in this letter might stimulate further experiments.

The required smallness of the ring traps forbids direct in-situ
imaging of the BECs. Therefore one has to employ time-of-flight
(TOF) imaging. We consider an experiment where initially two annular
BECs are created in two uncoupled ring trap from one single BEC.
\begin{figure}[htb]\center
\includegraphics[angle=0,width=8.5cm]{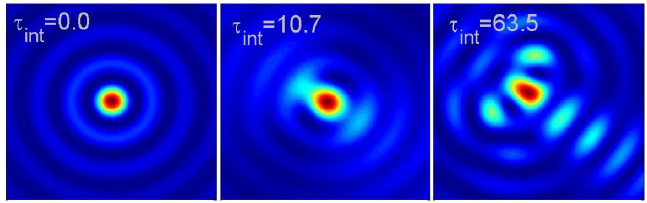}
\caption{(Color online) Simulated TOF images for several interaction
times. The parameters correspond to those of the top panel of figure
\ref{fig:selective_excitation}. The axis are in arbitrary units. For
an interaction time of $\tau_\text{int}=0.0$ only the $m=0$ mode
contributes to the momentum distribution yielding an image of the
squared zero-th Bessel function. At $\tau_\text{int}=10.7$ the
$m=\pm 1$ are occupied as well which gives rise to an angular
modulation proportional to $\sin^2(\phi)$ of the TOF image. For
later times several angular momentum modes are occupied. We show an
example for $\tau_\text{int}=63.5$. Here both annuli carry a net
angular momentum.  }
\label{fig:tof}
\end{figure}
Subsequently the barrier in $z$-direction is lowered such that a
certain coupling strength $\kappa$ is established. The system then
evolves at constant $\kappa$ for a certain time $\tau_\text{int}$
after which the trap is switched off. The TOF image is then taken
after free expansion of the cloud. Here we assume that only the wave
function of one annulus is imaged, i.e. the atoms in the second
annulus have to be removed \footnote{This can be achieved by first
separating the rings and discarding one of them}. The TOF method
yields an image of the momentum distribution of the BEC
\cite{Cozzini06,Lesanovsky06,Modugno06}. This is equivalent to the
squared modulus of the Fourier transform of the wave function
$\Psi(\mathbf{r})$. Using the angular momentum mode decomposition
$\alpha_m$ of the annulus which is to be probed and assuming that
$a_\rho,a_z \ll R$ we obtain
\begin{eqnarray}
\Psi(\mathbf{k})&\propto  & \sum_{m = \text{even}} (-1)^{m/2}
J_{|m|}(k R) \alpha_me^{i m \zeta} -\sum_{m = \text{odd}}
(-1)^{(m-1)/2}  J_{|m|}(k R) \alpha_me^{i m \zeta}\nonumber
\end{eqnarray}
with $J_n$ being the $n$-th Bessel function of the first kind,
$k=\sqrt{k_x^2+k_y^2}$ and $\zeta=\arctan(k_y/k_x)$. Hence imaging
$|\Psi(\mathbf{k})|^2$ allows the reconstruction of the $\alpha_m$
for a small number of contributing modes (for an example see
fig.$\,$\ref{fig:tof}). This allows an experimental study of the
instability regions simply by a analyzing TOF images.

In summary, in a system consisting of two ground state BECs in
coupled rings the occupation number of high angular momentum modes
grows exponentially for well-defined coupling strengths. For small
evolution times, a symmetric occupation of $\pm m$ modes takes place
in each BEC accompanied by ``ordinary" Josephson oscillations of the
relative particle number. For later times, angular momentum
Josephson oscillations spontaneously emerge. This novel type of
Josephson oscillations breaks the initial chiral symmetry of the
individual BECs.

This research project has been supported by a Marie Curie Transfer
of Knowledge Fellowship (IL) and a Marie Curie Excellence Grant (WK)
of the European Community's Sixth Framework Programme under the
contract numbers MTKD-CT-2004-014496 and MEXT-CT-2005-024854.

\end{document}